\documentclass[fleqn,usenatbib]{mnras}

\usepackage[T1]{fontenc}
\usepackage{ae,aecompl}

\usepackage{graphicx}	
\usepackage{amsmath}	
\usepackage{amssymb}	
\usepackage{gensymb}
\usepackage{mathrsfs}
\usepackage{txfonts, mathptmx}
\usepackage{dblfloatfix}
\usepackage{bm}
\usepackage{footmisc}
\usepackage{relsize}
\usepackage{booktabs}
\usepackage{hyperref}
\usepackage{units}

\newcommand{\ml}{ \Upsilon_{{\rm Disk}}}
\newcommand{\mlb}{ \Upsilon_{{\rm Bulge}}}
\newcommand{\kms}{\, \mathrm{km} \, \mathrm{s}^{-1}}

\newcommand{\ra}[1]{\renewcommand{\arraystretch}{#1}}

\newcommand{\vmax}{V_{\mathrm{max}}}
\newcommand{\vom}{V_{\mathrm{obs, \, max}}}

\newcommand{\Msun}{M_\odot}
\newcommand{\Mstar}{M_\star}
\newcommand{\Mhalo}{M_{{\rm halo}}}

\newcommand{\cvir}{c_{{\rm vir}}}
\newcommand{\rvir}{r_{{\rm vir}}}

\title[Dark Matter in Hydrodynamic Simulations]{Comparing  rotation curve observations to hydrodynamic $\Lambda$CDM simulations of galaxies. }
\author[A. B. Pace]{
Andrew B. Pace,$^{1}$\thanks{E-mail: apace@uci.edu }
\\
$^1$Center for Cosmology, Department of Physics and Astronomy, University of California, Irvine, CA 92697, USA
}

\date{Accepted XXX. Received YYY; in original form ZZZ}

\pubyear{2016}

\begin{document}
\label{firstpage}
\pagerange{\pageref{firstpage}--\pageref{lastpage}}
\maketitle

\begin{abstract}
The formation of the disk and feedback from supernova winds impacts the distribution of dark matter in galaxies.
Recently, \citet{Di_Cintio_2014_profile} characterized the halo response from baryonic processes in hydrodynamical simulations via a dependence on the ratio of stellar-to-halo mass ($\Mstar/\Mhalo$).
The (stellar) mass dependent halo profile links together the local and global properties of the halo (e.g. inner slope and $\Mhalo$) which allows for measurements of $\Mhalo$ without virial tracers.
We compile a large sample of rotation curves from the literature to test this halo profile.
We find that this halo profile can explain
rotation curve observations over a wide range of $\Mstar$.
However, the global results from our sample are inconsistent with a $\Lambda$ cold dark matter universe.
We do not find the expected correlation between the halo concentration and $\Mhalo$ and there is significantly larger scatter than expected.
Furthermore, a large portion of galaxies below $\Mstar\sim 10^{9}\,\Msun$  are found to be hosted by smaller halos than expectations from the abundance matching technique.
We find our results are robust to statistical priors and systematic effects such as  inclination angle, asymmetric drift correction, data source, and uncertainties in stellar mass-to-light ratios.
This suggests either a mischaracterization of the halo response due to  baryonic processes or additional non-standard dark matter physics.
\end{abstract}




\section{Introduction}

The $\Lambda$ cold dark matter ($\Lambda$CDM) paradigm successfully explains  the distribution of matter on large scales \citep[e.g. the 2dF Galaxy Redshift Survey;][]{Percival2001_2df} by postulating that dark matter halos are the sites of galaxy formation \citep{White1978, Blumenthal1984}.
There are several indirect, statistical methods utilized to associate galaxies and dark matter halos such as halo occupation distribution modeling \citep{Peacock2000, Benson2000, Berlind2002, Bullock2002, Kravtsov2004}, the conditional luminosity function \citep{Yang2003}, and the abundance matching technique \citep{, Vale2004, Vale2006, Conroy2006, Guo2010_abundance, Moster2010, Moster2013_shm, Behroozi2013_abundance}.
Abundance matching assumes that the cumulative number distributions of galaxies and halos are related in a monotonic manner; the most luminous galaxy is hosted by the most massive halo (within a given volume).
At  cluster scales ($\Mstar>10^{12}\Msun$),  X-ray mass measurements and virial scale tracers agree with abundance matching \citep[e.g.][]{Kravtsov2014}.
At lower masses and smaller scales there is a lack of virial  tracers to make direct halo mass ($\Mhalo$) measurements. 

Although the $\Lambda$CDM model can explain the large scale structure, there are several unresolved problems concerning the inferred structural properties of halos on galactic scales.
It is well established that the structural properties of dark matter halos in collision-less (dark matter-only) simulations  contain `cusps' ($\rho\sim r^{-1}$) in their central regions \citep[e.g.][]{NFW_1997, Bullock2001, Diemand2007, Maccio2007, Stadel2009, Zhao2009, Navarro2010, Klypin2011, Dutton2014, Klypin2016}.
In contrast, observations of galactic rotation curves prefer shallower central regions or `cores'    ($\rho\sim r^{0}$) \citep[e.g.][]{Flores1994, Moore1994, Salucci2000, Swaters2003, Gentile2004, Spekkens2005, Simon2005,  deblok2008, Oh2011, Adams2014,  Oh2015}. 
This discrepancy concerning the dark matter density inner slopes\footnote{In this work we refer to the inner slope as the log-slope at some small radii (1-2 kpc, see Equation~\ref{eq:logslope}).  In the literature, the inner slope commonly refers to as the log-slope at r=0.} is known as the `core-cusp' problem.

The `to-big-to-fail' (TBTF) problem concerns the over-abundance of massive, dense (sub-)halos in dark matter-only simulations compared with the abundance of observed 
(satellite-)galaxies 
\citep{BK_2011_TBTF, BK_2012_TBTF}.
It has been quantified by comparing the densities of  Milky Way satellite galaxies to their (over-dense) counterpart subhalos in Milky Way simulations \citep{BK_2011_TBTF, Purcell2012, Jiang2015} and with the lack of observed satellites with maximum dark matter circular velocities ($V_{{\rm max}}$) between $\sim30-55 \kms$ \citep{BK_2012_TBTF, Cautun2014, Jiang2015}.
Similar conclusions are reached for the M31 satellites \citep{Tollerud2014}, Local Group ($D\lesssim 1.2$ Mpc) objects  \citep{Kirby2014, Garrison-Kimmel2014}, and field galaxies  \citep{Ferrero2012, Miller2014, Papastergis2015, Klypin2015, Papastergis2015arXiv}.
In the field, the TBTF problem becomes apparent at $\vmax \lesssim25 \kms$ \citep{Papastergis2015} even when accounting for baryonic effects \citep{Papastergis2015arXiv}.
The inner regions of many observed galactic rotation curves contain less enclosed mass (both baryonic and dark matter) than is indicative of simulated galaxies with similar maximum observed circular velocity ($\vom$) \citep{McGaugh2007, Oman2015, Oman2016}.
These discrepancies indicate that it is unclear whether the  galaxies and halos are correctly matched in smaller halos.

The `core-cusp' and TBTF problems were identified with dark matter-only simulations and the inclusion of baryonic processes is a natural solution.
The formation of the galactic disk will steepen the central regions of the dark matter halo via adiabatic contraction \citep{Blumenthal1986, Ryden1987, Tissera1998, Gnedin2004, Gustafsson2006}.
Supernova feedback can create cores by driving stellar winds \citep{Navarro1996_feedback,Gnedin2002, Mo2004_feedback, Governato2010}, driving the bulk motion of the gas \citep{Mashchenko2006, Mashchenko2008}, or by creating fluctuations in the potential  \citep{Read2005, Pontzen2012}.
The transfer of angular momentum between infalling baryonic clumps and the dark matter halo could  produce shallower central regions (dynamical friction) \citep{El-Zant2001, Tonini2006, Romano_Diaz2008, Cole2011, Del_popolo_2014, Nipoti2015}.
The central densities could also be lowered from stellar feedback and tidal stripping (if a satellite) \citep{Pontzen2012, Zolotov2012, Brooks2013, Madau2014, Brooks2014, Arraki2014}.
The precise scales and effects of these solutions are under debate and other avenues have been considered.

There is a variety of non-standard dark matter physics that makes changes to the small scales without affecting large scales.
For example, self-interacting dark matter \citep{Spergel2000, Firmani2000} can lower the central densities and create dark matter cores \citep[e.g.][]{Rocha2013, Vogelsberger2014_sidm, Kaplinghat2014, Elbert2015, Fry2015}.
Recent work suggests that the TBTF problem could be alleviated on the dwarf galaxy scales  with warm dark matter \citep{Lovell2012, Abazajian2014, Lovell2014, Horiuchi2016}.
Late-decaying \citep{Wang2014},  scalar field \citep{Robles2015}, and late forming \citep{Agarwal2015} dark matter are also potential solutions to the TBTF problem.

Other authors have suggested that rotation curves do not accurately trace the potential.
Ignoring pressure support (generally accounted for with the application of the asymmetric drift correction) can bias the implied potential, especially for lower mass systems \citep{Rhee2004, Dutton2005, Valenzuela2007}.
There are several other systematics that have been discussed in the literature, for example: non-circular motions \citep{Swaters2003}, beam-smearing from the finite beam width in H{\sc i} observations \citep{van_den_bosch_2000, Swaters2003}, and axisymmetry issues \citep{Hayashi2004}. 
Rotation curve tests have been carried out by constructing realistic mock observations of hydrodynamic simulations  \citep{Rhee2004, Dutton2005, Valenzuela2007, kuzio_de_naray_2011, Oh2011_sim, Pineda2016}.
Several works have recovered the input halos and their slopes \citep{kuzio_de_naray_2011, Oh2011_sim} while others have inferred small cores in a cuspy halo when ignoring pressure support \citep{Pineda2016}. 
While addressing the validity of rotation curve measurements and examining exotic dark matter models are fruitful endeavors in this work we address  baryonic solutions.

Hydrodynamic simulations are required to test whether baryonic processes will alleviate small scale problems with the constraint that realistic galaxies are still formed (e.g. extremely efficient supernova feedback will remove  dark matter and create a core but may destroy the galaxy in the process).
There are several state-of-the-art hydrodynamic simulation projects \citep[e.g.][]{Stinson2013_magic,  Hopkins2014, Vogelsberger2014, Schaye2015_eagle, Wang2015} that utilize different star formation and feedback prescriptions  constructed with the aim to understand the formation and evolution of galaxies.
Hydrodynamic simulations are able to produce galaxies with realistic disks that lie on the Tully-Fisher relationship \citep{Robertson2004, Governato2007, Stinson2010, Piontek2011, Guedes2011, Christensen2012,Vogelsberger2014, Sales2016arXiv}.
They can create both bulgeless and realistic bulges \citep{Governato2010, Christensen2014, Snyder2015}, match observed colors \citep{Stinson2010, Sales2015}, match the size-luminosity relation \citep{Brooks2011}, and reproduce the stellar-to-halo mass relationship \citep{Guedes2011, Munshi2013, Hopkins2014, Di_Cintio_2014a, Wang2015}.
When stellar feedback is included, dark matter cores can be created \citep{Governato2010, Maccio2012_sim, Governato2012, Teyssier2013, Di_Cintio_2014a, Onorbe2015, Read2015, Chan2015_fireshapes, Tollet2016_shapes_nihao} but this is not ubiquitous as it depends on the particular feedback prescription as some projects lack cores \citep{Vogelsberger2014, Schaye2015_eagle}.
Hydrodynamic simulations of Milky Way-sized halos or Local Group-like objects have been found to alleviate the TBTF problem \citep{Zhu2016, Sawala2016, Wetzel2016}.

Simulations have shown that they are able to create dark matter cores but do these simulated cored galaxies correspond to cores in observed galaxies?
We address this by examining a (stellar) mass dependent halo profile.
In Section~\ref{section:2}, we introduce the halo profile, our rotation curve-fitting methodology, and the observational sample.
In Section~\ref{sec:results}, we show that the observed rotation curves are well reproduced and  compare the results to cosmological relationships.
In Section~\ref{sec:dis}, we discuss the potential systematics, the validity of the halo profile, and the implications of our findings.

\section{The Halo Response due to Baryonic Processes}
\label{section:2}

\begin{figure}
\begin{center}
\includegraphics[scale=.55]{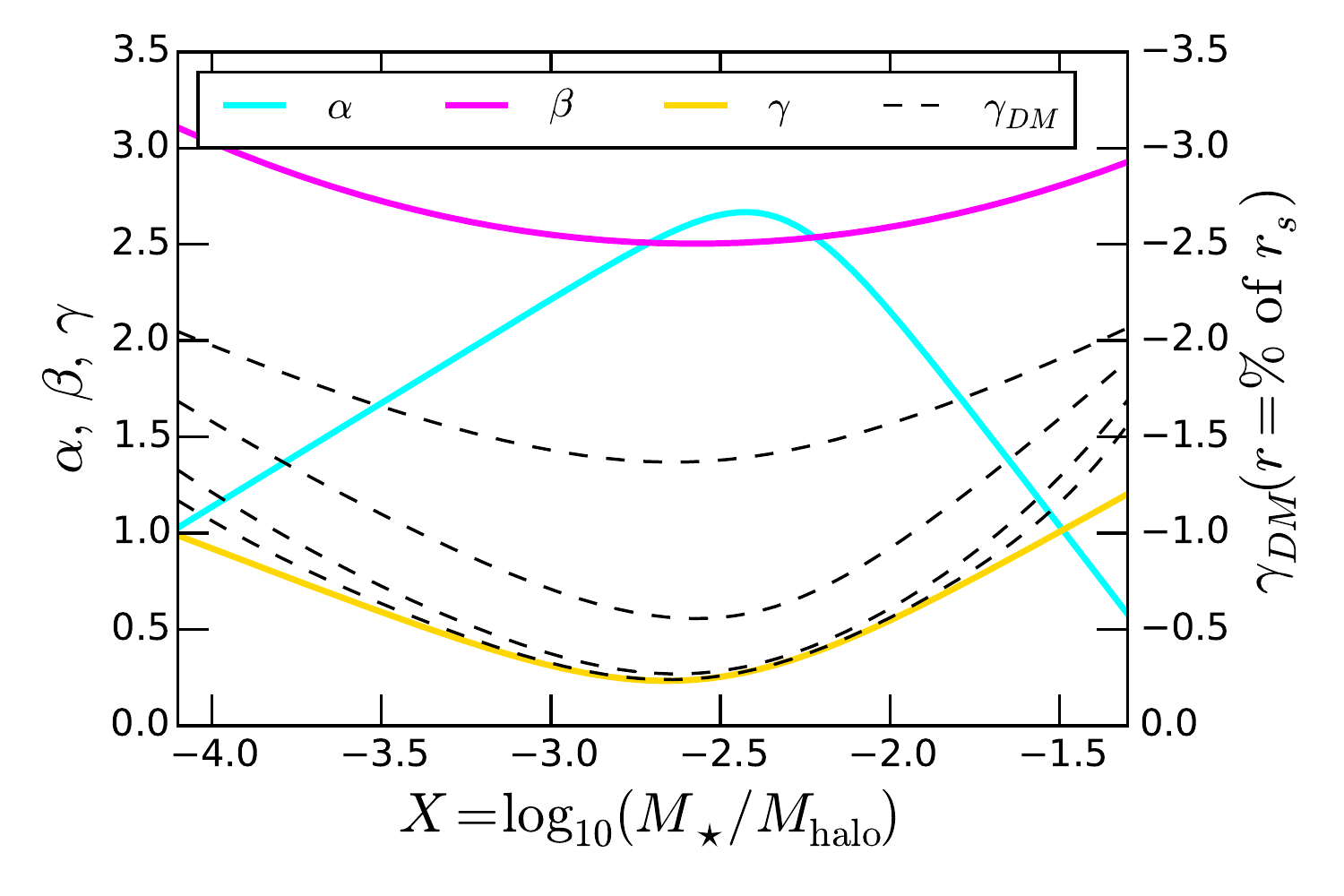}
\end{center}
\caption{
The dependence of the  inner ($\gamma$, gold), outer ($\beta$, magenta), and transition ($\alpha$, cyan) slope parameters  on $X\equiv\log_{10}{\left( \Mstar/\Mhalo\right)}$ for the DC14 profile (the values correspond to left y-axis; see Equation~\ref{eq:shape_parameters}). 
The dashed  black lines show log-slope dependence on $X$ for the DC14 profile  at different values of $r/r_s$ (the values are shown on the right y-axis; see Equation~\ref{eq:logslope}).
From bottom to top, the dashed lines correspond to $r/r_s=0.1, 0.2, 0.5, 1.0$  respectively.
}
\label{fig:dc14_slope}
\end{figure}

\citet[][hereafter DC14]{Di_Cintio_2014_profile} examined hydrodynamic simulations from the Making Galaxies In a Cosmological Context (MAGICC) project \citep{Stinson2013_magic} to determine how the dark matter distribution responds to baryonic processes and galaxy formation.
They determine the response by fitting $z\sim0$ halos with an $\alpha \beta \gamma$ density profile \citep{Jaffe1983, Hernquist1990, Merritt2006}:
\begin{equation} \label{eq:density}
\rho(r) = \rho_s \left(r/r_s\right)^{-\gamma} \left[1+\left(r/r_s\right)^{\alpha} \right]^{-(\beta-\gamma)/\alpha},
\end{equation}
\noindent and assuming  $\alpha$, $\beta$, and $\gamma$ are functions of the integrated star formation efficiency (parameterized by $X\equiv \log_{10}{\left( M_{\star}/\Mhalo \right)}$).
The functional forms with best fit values of $\alpha$, $\beta$, and $\gamma$ are (Equation 3 of DC14):
\begin{align}
\label{eq:shape_parameters}
\alpha(X) &= 2.94 - \log_{10}{\left[ \left( 10^{X+2.33} \right)^{-1.08} + \left( 10^{X+2.33} \right)^{2.29} \right]} \nonumber  \\
\beta(X) &=  4.23 + 1.34 X + 0.26 X^2 \nonumber  \\
\gamma(X) &= -0.06 + \log_{10}{\left[ \left( 10^{X+2.56} \right)^{-0.68} + \left( 10^{X+2.56} \right) \right]}.
\end{align}
\noindent The $X$ dependence shows the interplay between supernova feedback and adiabatic contraction. 
At high $X$ values the halo profile steepens due to the large baryonic content, while intermediate values have the shallowest inner slope due to efficient feedback.  
At low $X$ values the profile steepens again due to the inefficiency of star formation.

The log-slope of the $\alpha$, $\beta$, $\gamma$ profile is:
\begin{equation}
\gamma_{DM}(r) \equiv \left.\frac{ \mathrm{d}\ln{\rho}}{\mathrm{d}\ln{r^{\prime}}}\right\vert_{ r^{\prime }= r} = - \frac{\gamma + \beta \left( r/r_s\right)^{\alpha}}{1 + \left( r/r_s\right)^{\alpha}}=\gamma_{DM}(r; r_s, X),
\label{eq:logslope}
\end{equation}
\noindent where the last line follows for the DC14 profile.
In Figure~\ref{fig:dc14_slope}, we show $\alpha,\beta$, and $\gamma$ as a function of $X$.  Overlaid are (dotted-black) lines showing the $\gamma_{DM}$  dependence of a function of $X$ for fixed values of $r/r_s=0.1, 0.2, 0.5, 1.0$.
The minimum slope occurs at $X\sim-2.7$.  

We follow DC14 to define the halo concentration \footnote{This differs from the conventional definition but is identical for an NFW halo profile.} as:
$c_{{\rm vir}}\equiv r_{{\rm vir}}/r_{-2}$,
\noindent where $r_{-2}$ is $\gamma_{DM}(r=r_{-2})\equiv-2$. 
DC14 show that $\cvir$ is roughly equivalent between dark matter-only and hydrodynamic simulations if $X\lesssim-1.5$.
Similarly, the EAGLE project finds the dark matter-only and hydrodynamic $\cvir$ are consistent with one another \citep{Schaller2015}.

The MAGICC simulations fall within the following ranges: $-4.1<X<-1.3$, $2\times10^{5} \Msun<\Mstar<2.7\times10^{10} \Msun$, and $9.4\times10^{9} \Msun<\Mhalo<7.2\times10^{11} \Msun$ \citep[For exploration of the halo response for larger halos see][]{Dutton2015}.  
This profile has one additional parameter when compared to standard halo profiles and it encapsulates non-trivial baryonic processes.
The (stellar) mass-dependent halo profile ties together the inner properties ($\gamma_{DM}$) with the global  properties ($\Mhalo$).
We exploit this to infer $\Mhalo$ without tracers at the virial radius, $\rvir$.

This profile has already been used to model rotation curves for several galaxies \citep{Karukes2015, Repetto2015_dc} but not with a large sample.
It can potentially solve the TBTF problem in the Local Group \citep{Brook2015a} and explain the Tully-Fisher relation \citep{Brook2015b}.
When combined with scaling relations, it can potentially explain the scatter in rotation curve shapes \citep{Brook2015_variation}, and the mass discrepancy acceleration relation \citep{di_cintio_lelli_2016}.
\citet{Papastergis2015arXiv} test abundance matching using the velocity measured at the outermost radii  from the Arecibo legacy fast ALFA 21 cm survey.
Even after accounting for baryonic effects with the DC14 profile they find abundance matching breaks down in the field at low rotation velocities ($\vmax \lesssim 25 \kms$). 
The DC14 profile was created with fits to hydrodynamic simulations and has been utilized in statistical studies. 
We aim to remedy a weaknesses of the previous analysis  by testing this profile with a large observational sample of galactic rotation curves.

\subsection{Determining $\Mhalo$ and $\rho_s$}

In order to facilitate present and future comparisons an explicit definition of $\Mhalo$  is required.
We follow DC14 with updates to the latest Planck cosmology.
We define: $\Mhalo=\frac{4\pi}{3} \Delta \rho_{{\rm crit}} \rvir^3$, where $\Delta=18\pi^2+82x-39x^2=102.356$ ($x=\Omega_m-1$) \citep{Bryan1998} and $\rho_{{\rm crit}}=127.351\, {\rm \Msun \, kpc^{-3}}$ \citep[ $\Omega_m=0.3089$, ${\rm h}=0.6774$;][]{Planck2015}\footnote{For comparison, DC14 assumes a WMAP3 cosmology with $\Delta=93.6$, $\rho_{{\rm crit}}=147.896 \, {\rm \Msun \, kpc^{-3}}$.  We have verified that our results do not change between the WMAP3 and Planck cosmologies.}. 

To fully specify the halo profile, values of $r_s$, $\rho_s$, $\Mhalo$, and $\Mstar$ are required.
$\Mstar$ is defined as: $\Mstar=L_{x} \Upsilon_{{\rm photo}, \, x} \Upsilon_{{\rm kinematic}}$, where $L_x$ corresponds to the luminosity in the photometric band $x$.
$L_{x}$ and $\Upsilon_{{\rm photo}, \, x}$ are determined from the literature and $\Upsilon_{{\rm kinematic}}$ is treated as a free parameter.

The remaining parameters overdetermine the system; either $\rho_s$ or $\Mhalo$ can be eliminated.
We treat $\Mhalo$ as a free parameter and utilize the following prescription to determine $\rho_s$ (which is a modified form of the DC14 Appendix):
\begin{itemize}
  \item Determine $r_s$, $\Mhalo$, and $M_{\star}$ (via $\Upsilon_{{\rm kinematic}}$) from points in parameter space
  \item Determine $\rvir$ from $\Mhalo$
  \item Evaluate $M(\rvir)$ via the density profile: $M(\rvir) = 4 \pi \rho_s \int_0^{r_{{\rm vir}}} \left(r/r_s\right)^{-\gamma} \left[1+\left(r/r_s\right)^{\alpha} \right]^{-(\beta-\gamma)/\alpha} r^2 \mathrm{d}r$. 
  \item Solve for $\rho_s$ assuming $M(\rvir) = \Mhalo$.  
\end{itemize}
Note that treating $\rho_s$ as a free parameter is numerically impractical; as determining $\Mhalo$ from $r_s$ and $\rho_s$ involves solving an integral-differential equation.

\subsection{Priors and Parameter Estimation}

The  rotation curve includes contributions from the dark matter halo, gas disk, stellar disk, and  potential stellar bulge:
\begin{equation}
V_{\mathrm{tot}}^2 = V_{\mathrm{DM}}^2 + V_{\mathrm{Gas}}^2 +\ml V_{\mathrm{Disk}}^2 \,   \left( + \mlb V_{\mathrm{Bulge}}^2 \right).
\end{equation}
\noindent The baryonic components ($V_{\mathrm{Gas}}$, $V_{\mathrm{Disk}}$, $V_{\mathrm{Bulge}}$) are determined from the literature.  
We assume a factor of 1.4 when converting between the H{\sc i} and gas surface densities to account for primordial Helium and other elements.  
$V_{\mathrm{DM}}$ is determined from the halo circular velocity: $V_{\mathrm{DM}}^2(r)= G M_{\mathrm{DM}}(r)/r$.

To explore the parameter space and compute the Bayesian evidence for model selection, we utilize the Multi-Nested Sampling routine \citep{Feroz2008, Feroz2009}.  
Our likelihood is:
\begin{equation}
-2\ln{\mathscr{L}} \propto  \chi^2 = \sum_{i=1}^N \frac{\left[V_{i, \mathrm{obs}} - V_{\mathrm{tot}}(r_i) \right]^2}{\sigma_i^2}.
\end{equation}
We compute the Bayes' Factor for  model comparison tests\footnote{See \citet{Trotta2008} for a review of Bayesian model selection in astrophysics.}.
We generally do not compare the reduced $\chi^2$ as it only considers the best fit point and not the posterior distribution.
The Bayes' Factor is the ratio of the Bayesian evidence for two models\footnote{We refer to the logarithm of the Bayes factor as the Bayes factor in this manuscript.}:  $\ln{{\rm B}_{10}}=\ln{Z_{1}} - \ln{Z_{0}}$.  For $\ln{{\rm B}_{10}}>0$, model 1 is favored compared to model 0.  The significance is interpreted via Jefferys' scale; the $\ln{{\rm B}_{10}}$ ranges of  0-1, 1-2.5, 2.5-5, and $>5$ correspond to insignificant, mild, moderate, and significant evidence in favor of model 1 compared to model 0. 
The Bayes' Factor only considers comparisons of models and not overall goodness of fit.

Our prior distributions are:
\begin{itemize}
\item $r_s$: uniform in the range: $-1 < \log_{10}{\left(r_s/{\rm kpc}\right)} < 3$.
\item $\Mhalo$: uniform in the range: $5 < \log_{10}{\left(\Mhalo/\Msun\right)}< 14$.
\item $\Upsilon_{{\rm kinematic}}$: uniform in the range: $0.5 < \Upsilon_{{\rm kinematic}} < 2$.  The prior range is doubled for galaxies without  $\Upsilon_{{\rm photometric}}$ inferred from stellar population synthesis analysis.  For galaxies with a stellar bulge a second  $\Upsilon_{{\rm kinematic}}$ is included.
\item $\Mstar$ and $\Mhalo$ are kept within the range: $-4.1<X<-1.3$.
\item No cosmological priors are assumed between the halo parameters.
\end{itemize}

We assume the DC14 profile is valid throughout the entire $\Mhalo$ range.
We discuss enforcing the DC14 simulation limits in $\Mstar$ and $\Mhalo$ in Section~\ref{sec:final_sample}.
When available,  $\Upsilon_{{\rm photometric}}$ values are set by stellar population synthesis models \citep{Bell2001}. 

\subsection{Observational Sample}
\label{sec:data}

\begin{table*}
\caption{Galaxy sample and general properties.  See Section~\ref{sec:data} for explanation of columns. This table is available in its entirety in the online journal. The references are: a) \citet{Sorce2014};  b) \citet{Jacobs2009}; c) \citet{Dalcanton2009}; d) \citet{Chemin2006}; e) \citet{Puche1991_1}; f) \citet{Carignan1990}; g) \citet{Hlavacek-Larrondo2011};
}
\begin{tabular}{lll lll lll lll l }
\hline
Galaxy  &  D & Method &  Ref.  & $\langle i\rangle$  & $M_{{\rm Disk}}$  ($M_{{\rm Bulge}}$) & Sample &  ADC & $\Mstar$ Band & $\nicefrac{h_d}{r_d}$ & $\Upsilon$ & Q  & RC Ref. \\
(1) & (2) & (3) & (4) & (5) & (6) & (7) & (8) & (9) & (10) & (11) & (12) & (13)\\
\hline
NGC 24 & 9.60 & TF &  a & 64.0 & 9.47 & Misc & N  & I & 0 & Y & 1 &    d\\
NGC 45 & 6.64 & TRGB & b & 41.0 & 9.40 & Misc & N  & B & 0& Y &1 &    d\\
NGC 55 & 2.11 & TRGB & c & 77.0 & 9.67 & Misc & N  & B & 0 & N & 2 &   e \\
NGC 247 & 3.54 & TRGB &  c & 74.0 & 9.01 & Misc & N  & B & 0& Y & 1 &  f, g \\
\hline
\end{tabular}
\label{table}
\end{table*}

Our sample includes rotation curves from the following sources:  LITTLE THINGS \citep{Hunter2012, Oh2015}, THINGS \citep{Walter2008, deblok2008,Oh2008_things, Trachternach2008, Oh2011}, WHISP \citep{Swaters2002, Swaters2002_photo, Noordermeer2005_whisp3, Swaters2009}, the Ursa Major cluster \citep{Tully1996_uma1, Tully1997_uma2, Sanders1998_rc_uma_data, Trentham2001_uma3, Verheijen2001_rc_uma_data, Verheijen2001, Bottema2002, Bottema2002_more}, low surface brightness galaxies
 \citep{van_der_hulst_1993,deblok1996, McGaugh2001, deblok2002, Swaters2003, KDN_2006, Kuzio_de_Naray2008}, and a miscellaneous sample \citep{Begeman1987, Carignan1988, Jobin1990AJ,Lake1990AJ,Cote1991,Gonzalez_1991, Blais-Ouellette1999, van_Zee_1999, Weiner2001,Blais-Ouellette2001,Weldrake2003_ngc6822, Gentile2004, Gentile2007, Gentile2010,Elson2010, Kreckel2011, Frusciante2012,Lelli_2012_ugc,Fraternali2011,  Carignan2013,Elson2013, Corbelli2014, Lelli2014,Kam2015, Richards2015, Randriamampandry2015,Karachentsev2015, Bottema2015, Carignan1990, Puche1990_300, Puche1991_1, Puche1991_2, Chemin2006,  Hlavacek-Larrondo2011, Hlavacek-Larrondo_2011_ngc253,Westmeier2011,  Westmeier2013, Lucero_2015_ngc253}.
Galaxies with multiple rotation curve measurements are combined  in non-overlapping regions and higher resolution data is used in overlapping regions\footnote{Typically, optical H$\alpha$ is used in the inner regions and radio H{\sc i} measurements in the outer regions.}.

We define the rotation curve quality tag, Q (varying between 1-3 with 1=best), to tag systems that may have misestimated errors or systematics that indicate an untrustworthy rotation curve.
The quality decreases for galaxies containing the following: low kinematic inclination angles, $i<35\degree$, non-circular motions, disturbed velocity fields, asymmetries between the receding and ascending sides, or the presence of a star-burst phase.  
Q=1 systems contain none of these systematics, Q=2 systems contain 1-2 systematics, and Q=3 systems contain 3-4 systematics.  
In addition, the galaxies UGC 668 \citep[IC 1613;][]{Oh2015}, UGC 4305 \citep[DDO 50;][]{Oh2015}, and NGC 4736 \citep{deblok2008} are removed from the analysis.

The galaxy sample and properties are summarized in Table~\ref{table}. 
The columns denote: (1) galaxy name; (2) distance in Mpc;  (3) distance method; (4) distance reference; (5) average kinematic inclination angle, $\langle i\rangle$; (6) mass of the stellar disk in $\Msun$ (and potential stellar bulge);  (7) data source/survey;  (8) asymmetric drift correction (ADC); (9) photometric band utilized for $\Mstar$ measurements; (10) ratio of scale height to scale length; (11) stellar population synthesis model application for  $\Upsilon_{{\rm photometric}}$; (12) rotation curve quality tag; (13) rotation curve citation.
The distance methods are: Tully-Fisher (TF), tip of the red giant branch (TRGB), and Cepheid (Cep).
The asymmetric drift correction options are: application (Y), not applied (N), and note required (NR).
In the later case the effect was calculated and found to be sub-dominant. 
The ratio of scale height to scale length is denoted $\nicefrac{h_d}{r_d}$ and $\nicefrac{h_d}{r_d}=0$ denotes an infinitely-thin disk.
Rotation curve sources listed in parenthesis are unused.

Different photometric bands and methods  are utilized to determine $\Mstar$ and $\Upsilon_{{\rm photometric}}$.  
For example, THINGS and LITTLE THINGS utilize Spitzer Space telescope 3.6$\mu$m measurements and stellar population synthesis models to determine $\Mstar$.
The WHISP survey uses R-Band photometry and assumes $\Upsilon_{{\rm photometric}}=1$ for each galaxy.
For some galaxies, $\Upsilon_{{\rm photometric}}$ corresponds to the best fit value to the rotation curve \citep[e.g.][]{Cote2000, Gentile2004}.
In all cases we include a $\Upsilon_{{\rm kinematic}}$ as a free parameter.

\section{Results}
\label{sec:results}

\begin{figure*}
\begin{center}
\includegraphics[scale=.55]{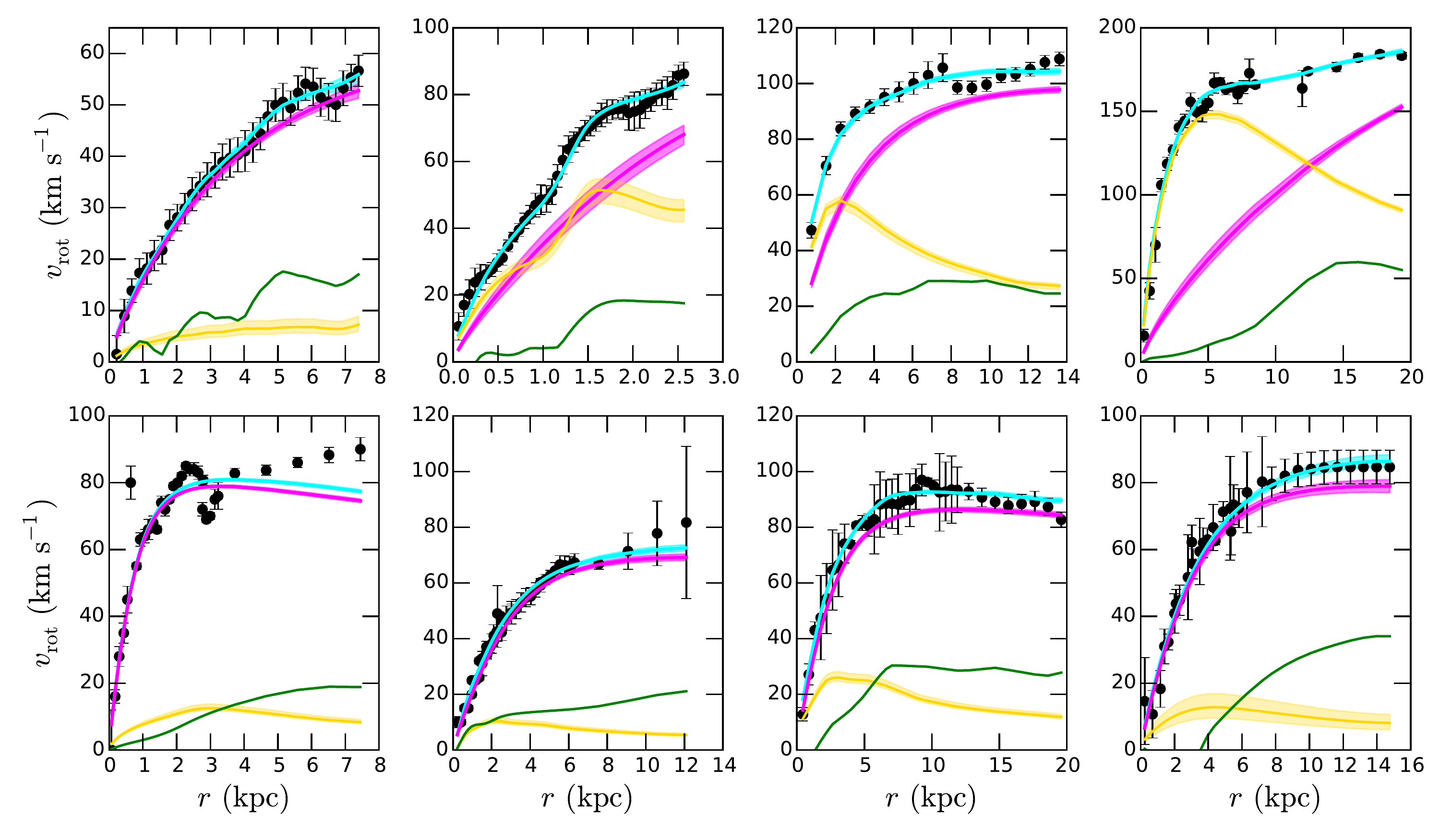}
\end{center}
\caption{
Examples of rotation curve fits with the DC14 (stellar) mass dependent halo profile.  Galaxies are  chosen from the $Q=1$ sample (see Section~\ref{sec:data}) to highlight the variety in the sample and both good and poor fits. The observational data is shown as black points with error bars.  The lines and shaded bands represent the following contributions: dark matter (magenta), stellar disk (gold),  gas disk (green), and  total fit (cyan).  The shaded bands correspond to the 68\% confidence interval (1-sigma region).
From left to right the galaxies are, {\bf top: } UGC 5918 \citep[LITTLE THINGS;][]{Oh2015}, NGC 2976 \citep[THINGS;][]{deblok2008}, NGC 4288 \citep[WHISP;][]{Swaters2009}, ESO 287-G15 \citep{Gentile2004}, {\bf bottom: } UGC 2259 \citep{Carignan1988, Blais-Ouellette2004}, NGC 3109 \citep{Jobin1990AJ, Blais-Ouellette2001, Carignan2013}, NGC 300 \citep{Puche1990_300, Westmeier2011, Hlavacek-Larrondo2011}, and F583-01 \citep{deblok1996, McGaugh2001, KDN_2006}.  
}
\label{fig:fits}
\end{figure*}

\begin{table*}
\ra{1.3}
\caption{Posteriors and model selection tests.  $\ln{{\rm B}}>0$ favors the DC14 model over the listed model. This table is available in its entirety in the online journal.
}
\begin{tabular}{lll lll lll lll lll l}
\hline
Galaxy & $\log_{10}{\left(r_s/{\rm kpc}\right)}$ & $\log_{10}{\left(\Mhalo/\Msun \right)}$ & $\Upsilon_{{\rm kinematic}}$ & $\log_{10}{\left(X\right)}$ & $\gamma_{DM}(1\, {\rm kpc})$ & $\gamma_{DM}(2\, {\rm kpc})$ & $\ln{{\rm B}_{{\rm PISO}}}$ & $\ln{{\rm B}_{{\rm NFW}}}$ & $\ln{{\rm B}_{{\rm Burkert}}}$ & $\chi_{r, \, DC14}^2$ & $\chi_{r, \, PISO}^2$\\
\hline
NGC 24  &  $0.96_{-0.09}^{+0.08}$  &  $11.50_{-0.08}^{+0.09}$  &  $1.25_{-0.28}^{+0.27}$  &  $-1.94_{-0.04}^{+0.04}$  &  $-0.62_{-0.04}^{+0.04}$  &  $-0.69_{-0.05}^{+0.05}$  &  -0.47  &  0.27  &  -0.43  &  0.30  &  0.26 \\
NGC 45  &  $0.67_{-0.02}^{+0.02}$  &  $11.17_{-0.01}^{+0.01}$  &  $0.72_{-0.08}^{+0.09}$  &  $-1.92_{-0.05}^{+0.05}$  &  $-0.71_{-0.05}^{+0.05}$  &  $-0.94_{-0.04}^{+0.04}$  &  39.90  &  -9.63  &  38.44  &  10.69  &  18.64 \\
NGC 55  &  $0.97_{-0.08}^{+0.16}$  &  $11.29_{-0.09}^{+0.21}$  &  $0.38_{-0.09}^{+0.17}$  &  $-2.06_{-0.07}^{+0.06}$  &  $-0.52_{-0.05}^{+0.05}$  &  $-0.58_{-0.06}^{+0.06}$  &  1.18  &  1.14  &  6.31  &  0.22  &  0.35 \\
NGC 247  &  $0.78_{-0.05}^{+0.03}$  &  $11.10_{-0.04}^{+0.03}$  &  $1.80_{-0.22}^{+0.14}$  &  $-1.84_{-0.03}^{+0.03}$  &  $-0.76_{-0.03}^{+0.03}$  &  $-0.92_{-0.04}^{+0.04}$  &  0.48  &  13.33  &  0.73  &  2.16  &  2.23 \\
\hline
\end{tabular}
\label{table:results}
\end{table*}

We apply the DC14 halo profile to our literature rotation curve sample and provide example fits  in Figure~\ref{fig:fits}.
The sample galaxies were chosen from the Q=1 subset to highlight the variety of rotation curves in the sample and to show examples of both good and poor fits.

The majority of the sample is well explained using the DC14 halo profile.
We quantify this by computing the reduced chi squared, $\chi_r^2$, which indicates good fits for most of the sample; $\chi_r^2<1$ for 76\% out of 177 galaxies.

We conduct comparisons between the DC14 profile and the Pseudo-Isothermal sphere\footnote{We also conducted fits with the Navarro-Frenk-White (NFW; \citeyear{NFW_1997}; $\alpha=1$, $\beta=3$, $\gamma=1$), and Burkert (\citeyear{Burkert_1995}) profiles. For most systems, the PISO provides better fits than the NFW and Burkert profiles, therefore we only conduct comparisons with the PISO profile in the main text. $\ln{{\rm B}}$ comparisons with the DC14 profile are listed in Table~\ref{table:results}.} (PISO; in terms of Equation~\ref{eq:density}, $\alpha=2$, $\beta=2$, $\gamma=0$).
The PISO profile is a commonly utilized `cored' halo profile in rotation curve analysis and generally provides good fits to rotation curves.
For our sample, the PISO profile provides a similar number of good fits with $\chi_{r, \, PISO}^2<1$ (72\%) and the median difference between the DC14 $\chi_r^2$ and PISO $\chi_r^2$ is $\langle \chi_{r, \, DC14}^2 - \chi_{r, \, PISO}^2 \rangle = 0.00_{-0.09}^{+0.07}$.
We turn to the Bayes' Factor which considers the entire posterior distribution, the size of the prior distribution, and number of parameters.  We the PISO profile we used a density scale, $\rho_s$, as a free parameter instead of $\Mhalo$.

Comparisons with the Bayes' Factor are similar; roughly half the sample has an indeterminate Bayes' Factor ($-1<\ln{{\rm B}}<1$, 52\%).
The reminder is divided between favoring the DC14 model ($\ln{{\rm B}}>1$, 20\%) and favoring the PISO model ($\ln{{\rm B}} < -1$, 29\%), although several systems significantly favor the PISO profile ($\ln{{\rm B}} < -5$, 7\%).

In Table~\ref{table:results}, we tabulate the median posterior values for $\log_{10}{\left(r_s/{\rm kpc}\right)}$, $\log_{10}{\left(\Mhalo/\Msun \right)}$, $\Upsilon_{{\rm kinematic}}$, $X$, $\gamma_{DM}(1 {\rm kpc})$, and $\gamma_{DM}(2 {\rm kpc})$. 
Additionally included, are the Bayes' Factors and $\chi_r^2$.

We reiterate that the $X$ dependence links together $\gamma_{DM}$ and $\Mhalo$ for the DC14 halo profile.
The physically motivated DC14 halo profile can explain rotation curve observations and is not disfavored compared to commonly utilized halo profiles.

\subsection{Multimodal Posteriors}
\label{sec:multimodal}

\begin{figure*}
\begin{center}
\includegraphics[scale=.45]{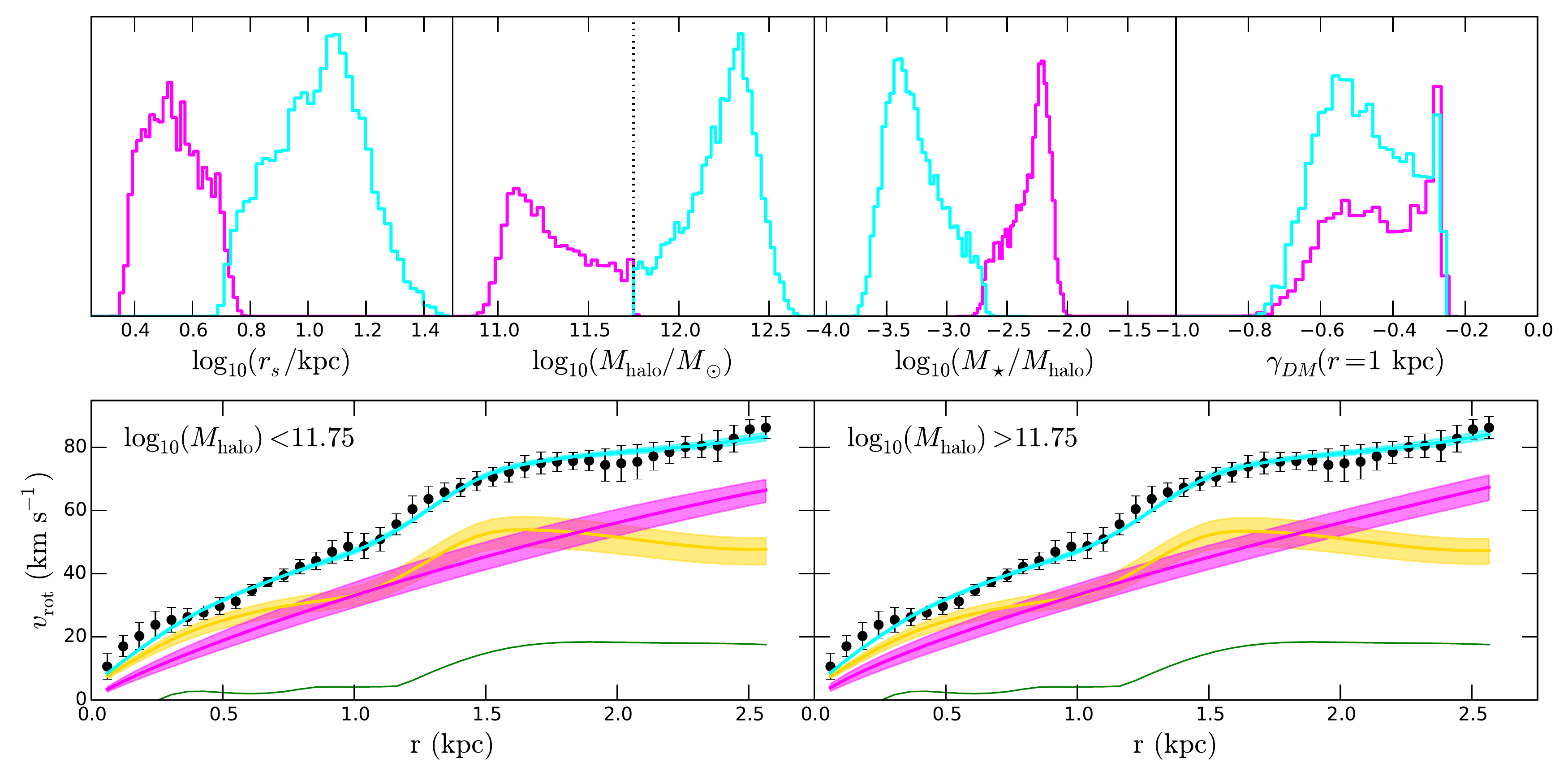}
\end{center}
\caption{Example of a galaxy with a multimodal posterior \citep[NGC 2976;][]{deblok2008}.
Without a well determined measurement of the flat component of the rotation curve the $\Mhalo$ posterior is multimodal.
{\bf Top: } Posterior distribution split into the two $\Mhalo$ modes.
The posteriors from left to right are: $\log_{10}{\left( r_s/{\rm kpc} \right)}$, $\log_{10}{\left(\Mhalo/\Msun\right)}$,  $\log_{10}{\left(\Mstar/\Mhalo\right)}$, and $\gamma_{DM}(r=1\, {\rm kpc})$. 
The two $\Mhalo$ modes are separated by applying a cut at $\log_{10}{\left(\Mhalo/\Msun\right)} =11.75$.
The smaller (magenta) and larger (cyan) modes contain 36\% and 64\% of the posterior respectively.
{\bf Bottom: }  Rotation curve fits corresponding to the two $\Mhalo$ modes.
The left (right) panel shows the fit for $\log_{10}{\left(\Mhalo/\Msun\right)} < 11.75$ ($>11.75$) (see Figure~\ref{fig:fits} for fit with the entire posterior).
The shaded bands represent 68\% confidence intervals for the subsets.
The lines and colors follow Figure~\ref{fig:fits}.
}
\label{fig:ngc2976}
\end{figure*}

Measurements of the central region of a galaxies rotation curve implies a value for the inner slope.
For most values of $\gamma_{DM}(r)$, there are two corresponding values of $X$ for a fixed value of $r_s$ (see Figure~\ref{fig:dc14_slope}).
This degeneracy is broken by measurements of the outer regions of the galaxy.
Due to the variety of data quality this measurement is not available in all systems.
Bi-modal $\Mhalo$ posteriors are inferred in many systems and there are additional degeneracies with $\Upsilon_{{\rm kinematic}}$ and $r_s$.

As an example, we consider high resolution H{\sc i} rotation curve of NGC 2976 from the THINGS survey \citep{deblok2008} where the flat component is not observed.
In Figure~\ref{fig:ngc2976}, we examine the posterior and rotation curve fit.
The upper panels display the posterior distributions  of $\log_{10}{\left(r_s/{\rm kpc} \right) }$, 
$\log_{10}{\left(\Mhalo/\Msun\right) }$, 
$X$, and $\gamma_{DM}(r=1\, {\rm kpc})$.
The posterior is separated at $\log_{10}{\left(\Mhalo/\Msun\right)} = 11.75$. 
Each $\Mhalo$ mode contains a corresponding mode in the $r_s$ and $X$ distributions, whereas the two modes have similar distributions for $\gamma_{DM}(r=1\,{\rm kpc})$.
In the lower panels, we show the rotation curve fits with the separated posterior (see Figure~\ref{fig:fits} for the full rotation curve).
Remarkably, the total circular velocity from each $\Mhalo$ mode is quite similar even though $\Mhalo$ differs by an order of magnitude.

We apply a similar analysis to all multimodal systems.
The posterior is separated at the minimum $\Mhalo$ value between the modes.  
In some galaxies, the separation is unclear.  
We denote these systems as poorly-separated and the former as well-separated.  

\subsection{Final Sample Selection}
\label{sec:final_sample}

\begin{figure*}
\begin{center}
\includegraphics[scale=.475]{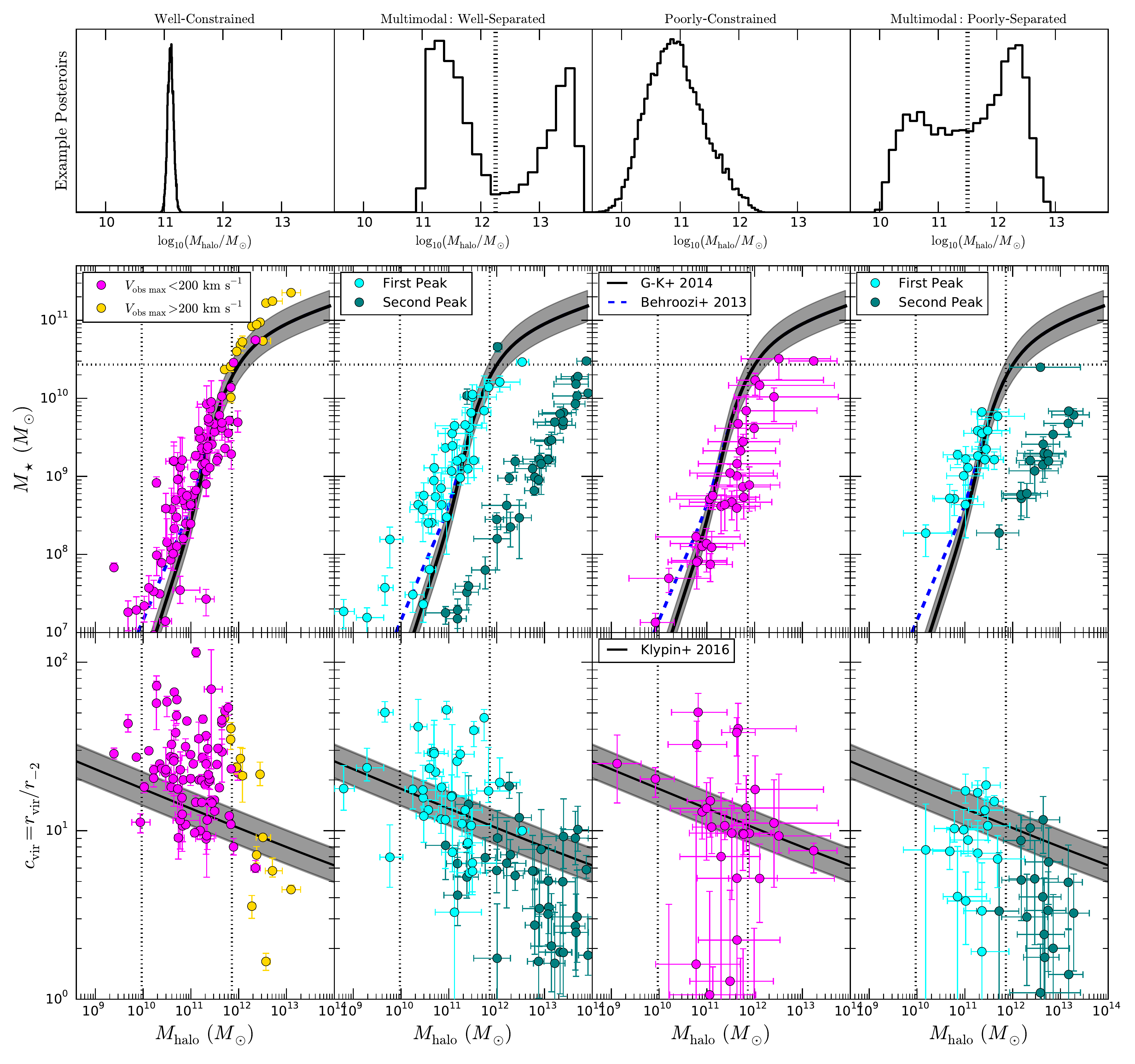}
\end{center}
\caption{
Visualization of the sample for $\Mhalo$ versus $\Mstar$ and $\cvir$ (See Figure~\ref{fig:mstar_mhalo} for the results with the final sample). 
The sample is split based on the distribution and number of modes in the $\Mhalo$ posterior.
The columns show the different divisions based on $\Mhalo$;
from {\bf left to right} the divisions are: single mode well-constrained, multimodal well-separated, single mode poorly-constrained, and multimodal poorly-separated.
{\bf Top: } example $\Mhalo$ posteriors from each division.
{\bf Middle: } $\Mhalo$ versus $\Mstar$. 
In the left-hand column, systems with $\vom>200\kms$ are shown as gold circles.
For multimodal systems, each mode is shown with different colored points.  The error bars correspond to 68\% confidence intervals within that mode.
Overlaid are abundance matching relationships \citet{Behroozi2013_abundance, Garrison-Kimmel2014_ELVIS} in blue and black respectively.
Dotted black lines show the DC14 simulation limits.
{\bf Bottom: } $\Mhalo$ versus $\cvir$.
Overlaid is the $\cvir-\Mhalo$ relationship from the Multidark simulations \citep{Klypin2016}.
The final sample consists of the well-constrained (left-hand panel) and the first mode of the well-separated systems (cyan points in the second panel from the left) with median $\Mhalo$ and $\Mstar$ values less than a factor of two from the upper limits of the DC14 simulations.
 }
\label{fig:explain_cuts}
\end{figure*}

We consider the $\Mhalo$ posteriors to construct a final sample to compare to cosmological relations.
We remove single mode systems with $\sigma_{\Mhalo}>0.4 \, ({\rm dex})$ and refer to them as poorly-constrained systems.
The remainder of the single mode posteriors are referred to as the well-constrained systems.

It is unclear whether the DC14 profile is valid outside of the simulation limits (e.g. at larger $\Mhalo$ feedback from active galactic nuclei becomes important) and a suitable prior is required to enforce the simulation limits.
An observationally motivated prior is to apply a cut at $\vom=200 \kms$.
This removes most but not all of the galaxies with large $\Mhalo$ and even removes systems well within the DC14 limits.
We therefore consider the $\Mhalo$ and $\Mstar$ posteriors for the cutoff; galaxies with median values of $\Mstar$ and $\Mhalo$ greater than a factor of two above the DC14 limits are excluded from the final sample.
We assume the DC14 profile is valid for galaxies with $\Mhalo$ smaller than the DC14 simulation limits (only 8 galaxies have median $\Mhalo$ below the simulation limits).

Determining which mode to consider in multimodal systems will affect the interpretation of our results.
We exclude all poorly-separated systems and consider the smaller mode of the well-separated posteriors (still considering the same cut in the median $\Mstar$ and $\Mhalo$ posteriors).
In most cases the second (larger) $\Mhalo$ mode is larger than the DC14 simulation limits and corresponds to an unrealistically large halo for the given galaxy (many of the larger modes have group or cluster $\Mhalo$). 
The results from the first $\Mhalo$ mode match the results for the well-constrained systems, providing circumstantial evidence that the larger $\Mhalo$ mode is unrealistic.
The final sample contains 119 galaxies after the application of these cuts.

We show the four $\Mhalo$ subsets visually in Figure~\ref{fig:explain_cuts}.
The columns show (from left to right): well-constrained, well-separated, poorly-constrained, and poorly-separated systems.
The top row shows an example posterior from each category. 
The middle and bottom rows show $\Mstar-\Mhalo$ and $\cvir-\Mhalo$ respectively.
In the well-constrained column, we show the systems with $\vom>200\kms$ as gold points; displaying the issue with a $\vom$ cut. 
Overlaid are $\Mstar-\Mhalo$ abundance matching relations \citep{Behroozi2013_abundance,Garrison-Kimmel2014_ELVIS} and $\cvir-\Mhalo$ relations from the MultiDark simulations \citep{Klypin2016}.
The multimodal systems separate in both the $\Mstar-\Mhalo$ and $\cvir-\Mhalo$ space.
The final sample consists of the well-constrained galaxies (left-hand column) and the first mode of the well-separated multimodal systems (cyans points; second column from the left).

\subsection{Cosmological Relations}
\label{section:SHM}

\begin{figure}
\begin{center}
\includegraphics[scale=.425]{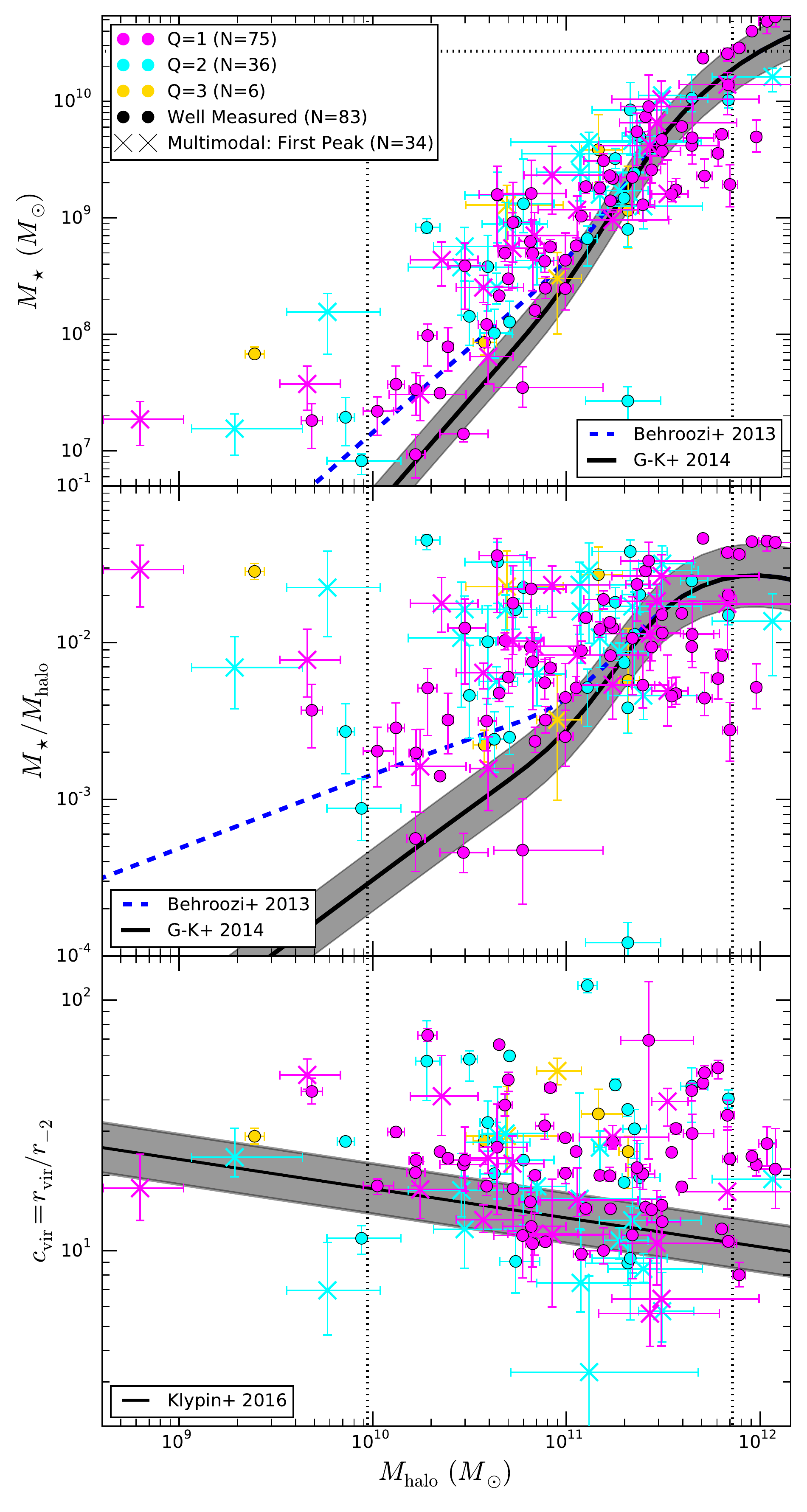}
\end{center}
\caption{Cosmological relations inferred from galactic rotation curve fits with the DC14 profile  (see Section~\ref{sec:final_sample} for an explanation of our sample selection). 
We denote potential systematics with the quality tag and separate them by color (see Section~\ref{sec:data}; Q=1 denotes trustworthy rotation curves).  
The symbols denote whether the  $\Mhalo$ posterior is multimodal. 
For multimodal systems we consider the first mode and errorbars denote 68\% confidence intervals within the first mode's posterior (See Figure~\ref{fig:ngc2976}, Section~\ref{sec:multimodal}).
{\bf Top: } $\Mstar$ vs $\Mhalo$. 
Overlaid are $\Mstar-\Mhalo$ relations derived from the abundance matching technique \citep[][are shown as blue-dashed and black-solid lines respectively]{Behroozi2013_abundance, Garrison-Kimmel2014_ELVIS}.
We assume 0.2 dex of spread in the $\Mstar-\Mhalo$ relations \citep{Behroozi2013_abundance, Reddick2013}.
The dotted lines show the range of simulations in DC14.
{\bf Middle: } $\Mstar/\Mhalo$ versus $\Mhalo$.  
The differences between abundance matching and the inferred halos are emphasized in this space.
{\bf Bottom: } $\cvir$ versus $\Mhalo$.
The black solid lines show the expected $\cvir-\Mhalo$ relationship from the MultiDark simulations \citep{Klypin2016}.
 }
\label{fig:mstar_mhalo}
\end{figure}

In Figure~\ref{fig:mstar_mhalo} we show the derived cosmological relations from the final sample compared to results from the literature.
The top panel shows the $\Mstar-\Mhalo$ relationship overlaid with the relationship derived from abundance matching \citep{Behroozi2013_abundance, Garrison-Kimmel2014_ELVIS}.
The galaxies are colored according to their quality tag and the symbols denote multimodal systems.

We focus our abundance matching comparison to two recent works focusing on different mass regimes \citep{Behroozi2013_abundance, Garrison-Kimmel2014_ELVIS}.
The first, \citet{Behroozi2013_abundance}, is constructed with large volume observations of the stellar mass function, cosmic star formation rate, and specific star formation rate and  is complete to $\Mstar \sim 10^{8.5} \, \Msun$ ($\Mhalo \sim 10^{10.9} \, \Msun$). 
The later \citep{Garrison-Kimmel2014_ELVIS}, use the local group (defined as galaxies within $\sim1.2$ Mpc) stellar mass function to push the completeness to $\Mstar\sim10^5\Msun$.
They tie their relation to \citet{Behroozi2013_abundance} at large masses and find a steeper faint end slope \citep[see also][]{Brook2014_SHM, Shea_GK_2016}.
We assume a spread of 0.2 dex \citep{Behroozi2013_abundance, Reddick2013} in the $\Mstar-\Mhalo$ relation and display this relative to the \citet{Garrison-Kimmel2014_ELVIS} relationship.
For $\Mstar > 10^9 \Msun$, we observe scatter relative to the abundance matching relationships.
Below this, the galaxies preferentially lie in smaller halos than expected from abundance matching.
Our results strongly disagree with the local group stellar mass function.

The middle panel displays $X=\Mstar/\Mhalo$ versus $\Mhalo$.  
Many of the systems with low $\Mhalo$ have significantly larger $\Mstar$ than expected.
For a given $\Mstar$, many galaxies are hosted by significantly smaller halos than expected.  
Galaxies produced in hydrodynamic simulations (including the MAGICC project) are found to match the $\Mstar-\Mhalo$ relationships within the regime of masses we are considering \citep{Munshi2013, Hopkins2014, Di_Cintio_2014a, Wang2015}.
The $\Mstar-\Mhalo$  relationship from \citet{Behroozi2013_abundance, Garrison-Kimmel2014_ELVIS} combined with the DC14 profile predicts a dark matter inner slope.
This prediction is at odds with the dark matter cores and inner slopes inferred  from rotation curve observations.
At a given $\Mhalo$, the dark matter inner slopes in the DC14 simulations do not correspond to the dark matter inner slopes in galaxy observations.

In the lower panel, we display $\cvir$ versus $\Mhalo$.
Overlaid is the $\cvir-\Mhalo$ relationship from the MultiDark simulations \citep{Klypin2016}.  
We observe much higher $\cvir$ than cosmological simulations.
$\cvir$ is not expected to change between hydrodynamic and dark matter-only simulations \citep[DC14;][]{Schaller2015}.
We do not expect the baryon response to affect the $\cvir-\Mhalo$ relation.
The observed $\cvir-\Mhalo$ relation contains significantly larger scatter than the relationship found in dark matter-only simulations.

Although we can explain rotation curve observations, we do not recover the $\Mstar-\Mhalo$ and $\cvir-\Mhalo$ cosmological relationships.  
In Appendix~\ref{app:deviation}, we compare the deviations between the $\Mstar-\Mhalo$ and $\cvir-\Mhalo$ relationship. 
We uncover no trends in the deviations.
Based on the above discussion, these relationships are not expected to differ when changing from dark matter-only to hydrodynamic simulations.

\section{Discussion}
\label{sec:dis}
\subsection{Stellar Mass and other Systematics}

\begin{figure*}
\begin{center}
\includegraphics[scale=.4]{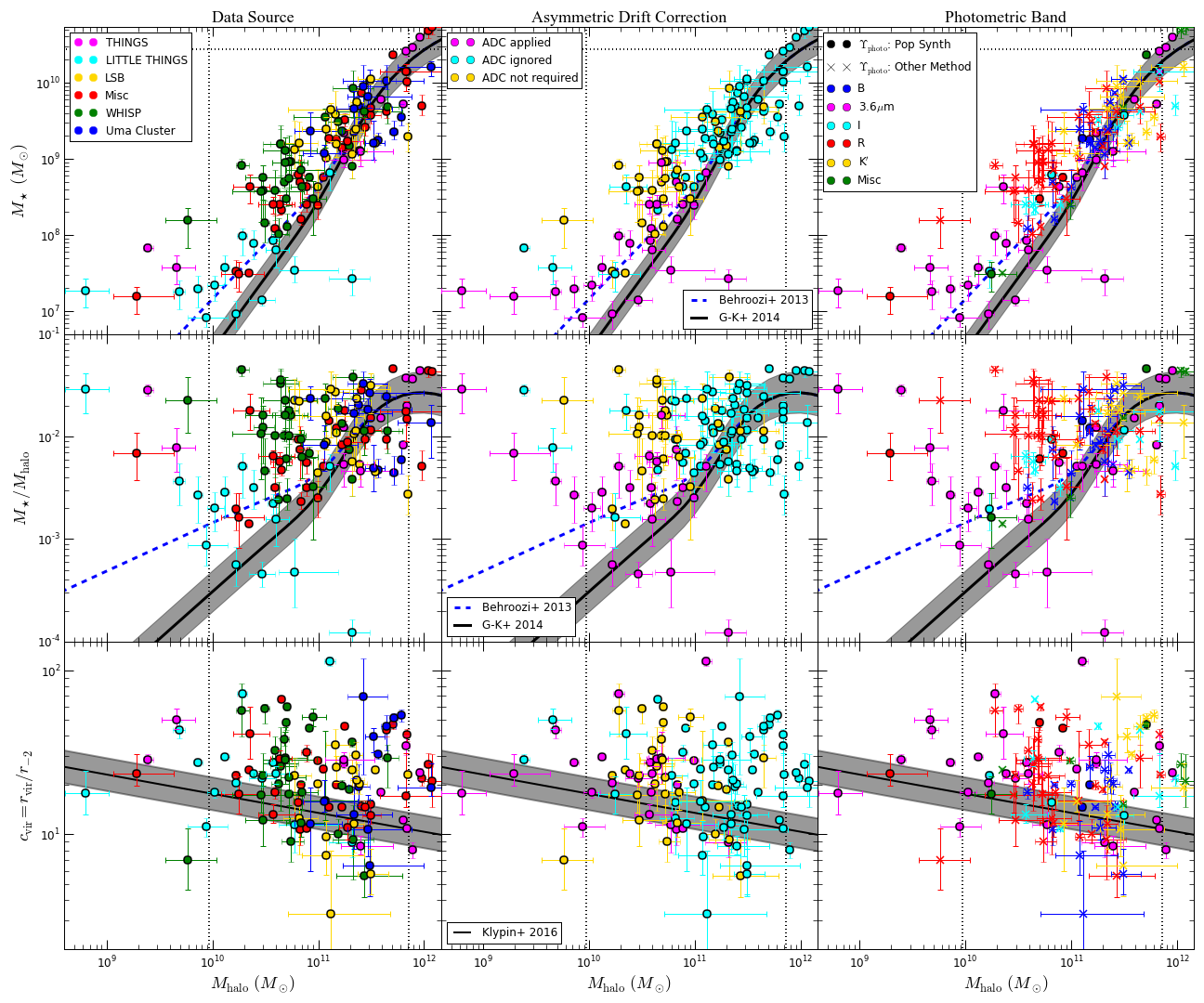}
\end{center}
\caption{$\Mstar-\Mhalo$, $\Mstar/\Mhalo-\Mhalo$, and $\cvir-\Mhalo$ relations for  subsets of the sample (See Figure~\ref{fig:mstar_mhalo}).
The subsets are: data source (left-hand column), asymmetric drift application (middle column), and $\Mstar$ photometric band (right-hand column). 
The data sources are (color and citation): THINGS \citep[magenta;][]{deblok2008, Oh2011}, LITTLE THINGS \citep[cyan;][]{Oh2015}, low surface brightness galaxies (LSB; gold; see Section~\ref{sec:data} for citations), miscellaneous (Misc; red; see Section~\ref{sec:data} for citations), WHISP \citep[green;][]{Swaters2009}, and Ursa Major cluster \citep[Uma; blue;][]{Sanders1998_rc_uma_data, Verheijen2001_rc_uma_data}.
The the asymmetric drift correction (ADC) subsets are:  application  (magenta), disregarded (cyan), and calculated but too small to affect the results (gold).
The photometric bands are: B (blue), Spizter space telescope 3.6$\mu$m (magenta), I (cyan), R (red), K$^{\prime}$ (gold), and Misc/other (green).
Galaxies with (without)  $\Upsilon_{{\rm photometric}}$ values calculated from stellar population synthesis models are shown as circles (x's).
The overlaid relationships follow Figure~\ref{fig:mstar_mhalo}.
}
\label{fig:systematics}
\end{figure*}

A data sample from the literature is heterogeneous and systematic biases may be introduced.
We address potential systematics by dividing the sample based on the application of different methods in interpreting the rotation curve.
The first is through the quality tag; galaxies with indications that the rotation curve may not trace the underlying potential are denoted with a higher tag (see Section~\ref{sec:data}).
At low inclination angles ($i<35\degree$) small changes to the inclination will result in large differences in the measured circular velocity.
A galaxy in a starburst or post-starburst phase will contain gas out of equilibrium resulting in rotation curves that may not match the true circular velocity \citep{Lelli2014, Read2016}.
Disturbed velocity fields, lopsided gas distributions, and asymmetries between the receding and ascending sides may be signs of recent star formation indicating additional uncertainties unaccounted for in the standard measurement errors \citep{Lelli2014_2}.
We separate the sub-samples by color in Figure~\ref{fig:mstar_mhalo}.  Our results are robust to the  removal of higher Q systems (see Figure~\ref{fig:mstar_mhalo_q_1}).
Furthermore, we find that our results are robust to  pressure support, uncertainties in $\Mstar$, and the combination of different surveys and data sources, as we discuss below in detail.

Pressure support in low mass systems ($\vom<75 \,\kms$) may lead to incorrect inferences of the circular velocity; the asymmetric drift correction is commonly used to correct for pressure support \citep{ Dalcanton2010}.
Roughly half of the sample either has the asymmetric drift correction applied (21\%) or pressure support has been determined and is to small to affect the rotation curve (26\%) \citep[e.g.][]{Weldrake2003_ngc6822, Swaters2009, Karachentsev2015}.
For low mass systems ($\vom<75 \,\kms$), 49\% of the systems have the asymmetric drift correction and for 33\% of the systems it is not required.

Incorrect measurements of $\Mstar$ will change the implied effect baryons have to the halo.
To address this we first  update the distance\footnote{The distances for the THINGS and LITTLE THINGS surveys are not changed.} (and therefore luminosity)  based on more precise measurements from the Tully-Fisher relation \citep{Tully2008, Sorce2014}, the  tip of the red-giant branch  \citep{Jacobs2009, Dalcanton2009}, or Cepheid variable star measurements.

Infrared photometry traces the old stellar population and is less affected by intergalactic dust \citep{Walter2007}.
Ideally, Spitzer 3.6${\rm \mu}$m would be utilized \citep[e.g.][]{deblok2008, Oh2015}, but it is not available for all galaxies. 
We separate the sample by the photometric band utilized to derive the stellar surface density and stellar luminosity.  
This subdivision is similar to a separation on data source but the miscellaneous portion contains a wide variety of photometric bands.
The primary bands utilized are: Spitzer 3.6${\rm \mu}$m (24\%), B (15\%), I (10 \%), K$^{\prime}$ (12\%), and R (32\%).
When available, $\Upsilon_{{\rm photometric}}$ is pinned to stellar population synthesis models (34\% of our sample).

The priors on   $\Upsilon_{{\rm kinematic}}$ varied by a factor of 2 or 4 depending on whether   $\Upsilon_{{\rm photometric}}$ was determined by stellar population synthesis models.
We explore using a much wider (and unrealistic) prior range, $0.05 < \Upsilon_{{\rm kinematic }} < 10.0$.
Our general results are robust to the larger prior but the results for individual galaxies are not. 
In some systems the  posterior in $\Mhalo$ becomes multimodal or  significantly increases in size due to degeneracies with  $\Upsilon_{{\rm kinematic}}$. 
We reconstruct Figures~\ref{fig:explain_cuts} and~\ref{fig:mstar_mhalo} with the  results from the larger prior range in Appendix~\ref{appendix}. 

The sample contains stellar disk rotation curves constructed with both the thin disk approximation and a non-zero scale height.
The  common ratios of scale height-to-scale length ($\nicefrac{h_d}{r_d}$) are: $\nicefrac{h_d}{r_d}=0,\nicefrac{1}{5}, \nicefrac{1}{6}$.
An assumption of $h_d=r_d/5$ is valid for many disk-dominated galaxies \citep{van_der_Kruit_1981, Kregel2002}.
This difference has a small effect but changes in the baryonic circular velocity will be reflected in the inferred dark matter halo.
Most of the sample contains a non-zero scale height (74\%) while the reminder assumes a thin disk.

Our results are robust to: pressure support (asymmetric drift correction), uncertainties in $\Upsilon$, the photometric band utilized for the $\Mstar$ measurement, the stellar disk circular velocity calculation, and the combination of different surveys.
In Figure~\ref{fig:systematics}, we reconstruct Figure~\ref{fig:mstar_mhalo} with the sample split by the data source (left-hand column), asymmetric drift correction application (middle cloumn), and photometric band utilized for $\Mstar$ measurements (right-hand column).  
Each subset has large scatter compared to the $\Mstar-\Mhalo$ and $\cvir-\Mhalo$ relations.
None of the subsets have significant offsets from the main sample; our results are not driven by a particular data source or photometric band.
It is unlikely that  observational systematics account for our results.

There are several improvements that can be made to this work.
Measurements of $\Mstar$ and the stellar surface density in a single consistent photomteric band, ideally infrared photometry, would decrease uncertainties in $\Mstar$ measurements and the systematics in combining different data sources. 
There are several methods for rotation curve construction that our sample contains \citep[e.g.][]{Sancisi1979, Begeman1987,van_der_hulst_1992_gipsy,deblok1997, Gentile2004, Spekkens2007, Oh2008_things, Oh2011, di_teodoro_2015}.
Construction of the rotation curve from the data cubes in a uniform manner would similarly reduce systematics.
Application of the asymmetric drift correction for all galaxies, especially for systems with $\vom < 75
\kms$, would reduce the uncertainties from pressure support \citep{Dalcanton2010}.

\subsection{Baryonic Halo Profiles}

The (stellar) mass-dependent profile of DC14 is thus far the only work that has characterized the response of  the \textit{entire} halo profile due to baryonic processes. 
It is well established that hydrodynamic simulations with stellar feedback can create dark matter cores and the focus has been on determining the size of the dark matter core or how the dark matter inner slope scales \citep{Governato2010, Maccio2012_sim, Governato2012, Teyssier2013, Di_Cintio_2014a, Onorbe2015, Read2015, Chan2015_fireshapes, Tollet2016_shapes_nihao}.
For example, \citet{Governato2012}  quantified the halo response of the dark matter inner slope to be: $\rho_{DM}\propto r^{\alpha};\,\,\alpha=-0.5+0.35 \times \log{10}{\left(\Mstar/10^{8}\Msun \right)}$.

Recent work has focused  on the dependence of the inner slope with  $X$  \citep{Chan2015_fireshapes, Tollet2016_shapes_nihao}.
The NIHAO suite ($\sim70$ simulations) contains an updated star formation and feedback prescription from the MAGICC simulations and finds a dependence that agrees with the DC14 profile \citep{Tollet2016_shapes_nihao}.
The FIRE project ($9$ simulations) uses an independent star formation and feedback prescription  with a pressure-independent smoothed particle hydrodynamics code \citep{Hopkins2014} and  finds the inner slope to have a different $X$ dependence \citep[See Figure 4 of ][]{Chan2015_fireshapes}.
The location of the minimum inner slope is the same in both works but \citet{Chan2015_fireshapes} has a steeper slope at small $X$ and shallower slope at large $X$ compared to \citet{Tollet2016_shapes_nihao}.
In both cases, observed galaxies with dark matter cores will be driven to $X\sim-2.7$ regardless of $\Mhalo$ or $\vom$ if the halo profile is of this form.
With such a small sample it is unclear how significant the discrepancy is and further work is required.

We define the core radius as\footnote{The core radius is commonly defined from the logarithmic slope or  density.  The density definition of $r_c$ is: $\rho(r_c)/\rho(0) \equiv \nicefrac{1}{2}$.
We favor the slope definition since the density is not finite at r=0.    For the PISO profile the two definitions are equivalent and they agree to within 20\% for the Burkert profile.}: $r_c = r_{-1}$, then $r_c=r_s \left(\frac{1-\gamma}{\beta-1} \right)^{\nicefrac{1}{\alpha}}$.
With this definition, we find the maximum $r_c$ at  $X\sim -2.7$ for fixed $r_s$.   
For low mass galaxies that favor large cores, $\Mhalo$ will be driven towards $X\sim -2.7$ which increases $\cvir$ for fixed $r_s$. 
This drives the galaxies away from cosmological relations.

The core radii correlates with the stellar radial scale in hydrodynamic simulations with stellar and supernova feedback, imprinting an additional radial scale in the halo \citep{Onorbe2015, Read2015}.
To fully capture the halo response a density profile with a second radial scale may be required \citep{Read2015}.
Observations find that the  stellar disk size scales with $\Mstar$ \citep[$r_d \propto \Mstar^{\alpha}$; e.g.][]{Hunter2006, Courteau2007, Fathi2010}, implying the $X$ dependence in the DC14 profile may already include the dependence of the stellar radial scale.
Exploring different functional forms may be a fruitful endeavor.
It is possible that the MAGICC simulations have not fully captured the halo response and a refined (stellar) mass-dependent profile can alleviate the tension observed between the rotation curve fits and cosmological relationships.
Current results from hydrodynamic simulations do not suggest strong deviations from the DC14 profile.

The breakdown of the cosmological scaling relations is 
indicative of the failure to solve the TBTF and diversity problems.
The observed inner slopes can be explained with baryonic physics but will not simultaneously match cosmological relationships.
Standard cosmological relationships breaking down at low $\Mhalo$ may be pointing towards unaccounted  effects and  additional physics in the dark sector may help reconcile the tension.
For example, warm \citep{Lovell2014}, self-interacting \citep{Rocha2013, Kaplinghat2014}, or scalar field \citep{Robles2013, Schive2014} dark matter can imprint another radial scale in the dark matter halo without  affecting large-scale structure.

We have conducted tests of the (stellar) mass dependent halo profile from DC14 with rotation curves from the literature.
The (stellar) mass dependent profile can explain rotation curve observations (i.e. solve the `core-cusp' problem) but will not simultaneously reproduce the cosmological $\Mstar-\Mhalo$ or $\cvir-\Mhalo$ relationships.
Directly modeling  rotation curves with halo profiles set by hydrodynamic simulations is a fruitful method to test the dark matter response to baryonic processes in hydrodynamic simulations.

\section*{Acknowledgements}

We thank Manoj Kaplinghat for helpful discussions and comments on the manuscript.
We thank James Bullock, Sheldon Campbell, Sean Tulin, and Hai-bo Yu for helpful discussions.
ABP is supported by a GAANN fellowship.
We thank the following for kindly sharing rotation curve data: Erwin de Blok, Roelof Bottema, Jayaram Chengalur,  St\'{e}phanie C\^{o}t\'{e}, Edward Elson, Noemi Frusciante,  Gianfranco Gentile, Igor Karachentsev, Kathryn Kreckel, Federico Lelli, Stacy McGaugh, Se-Heon Oh, Toky Randriamampandry, Emily Richards, Rob Swatters, and Ben Weiner.

Databases and software: This research has made use of the NASA/IPAC Extragalactic Database (NED) which is operated by the Jet Propulsion Laboratory, California Institute of Technology, under contract with the National Aeronautics and Space Administration and the HyperLeda database\footnote{\url{http://leda.univ-lyon1.fr}} \citep{Makarov2014}.
Python packages: \texttt{Astropy}\footnote{\url{http://www.astropy.org}} \citep{astropy2013},  \texttt{NumPy} \citep{numpy}, \texttt{iPython} \citep{ipython}, \texttt{SciPy} \citep{scipy}, and \texttt{matplotlib} \citep{matplotlib}.



\bibliographystyle{mnras}
\bibliography{main_bib_file} 

\appendix

\section{High Quality Rotation Curves}

\begin{figure}
\begin{center}
\includegraphics[scale=.45]{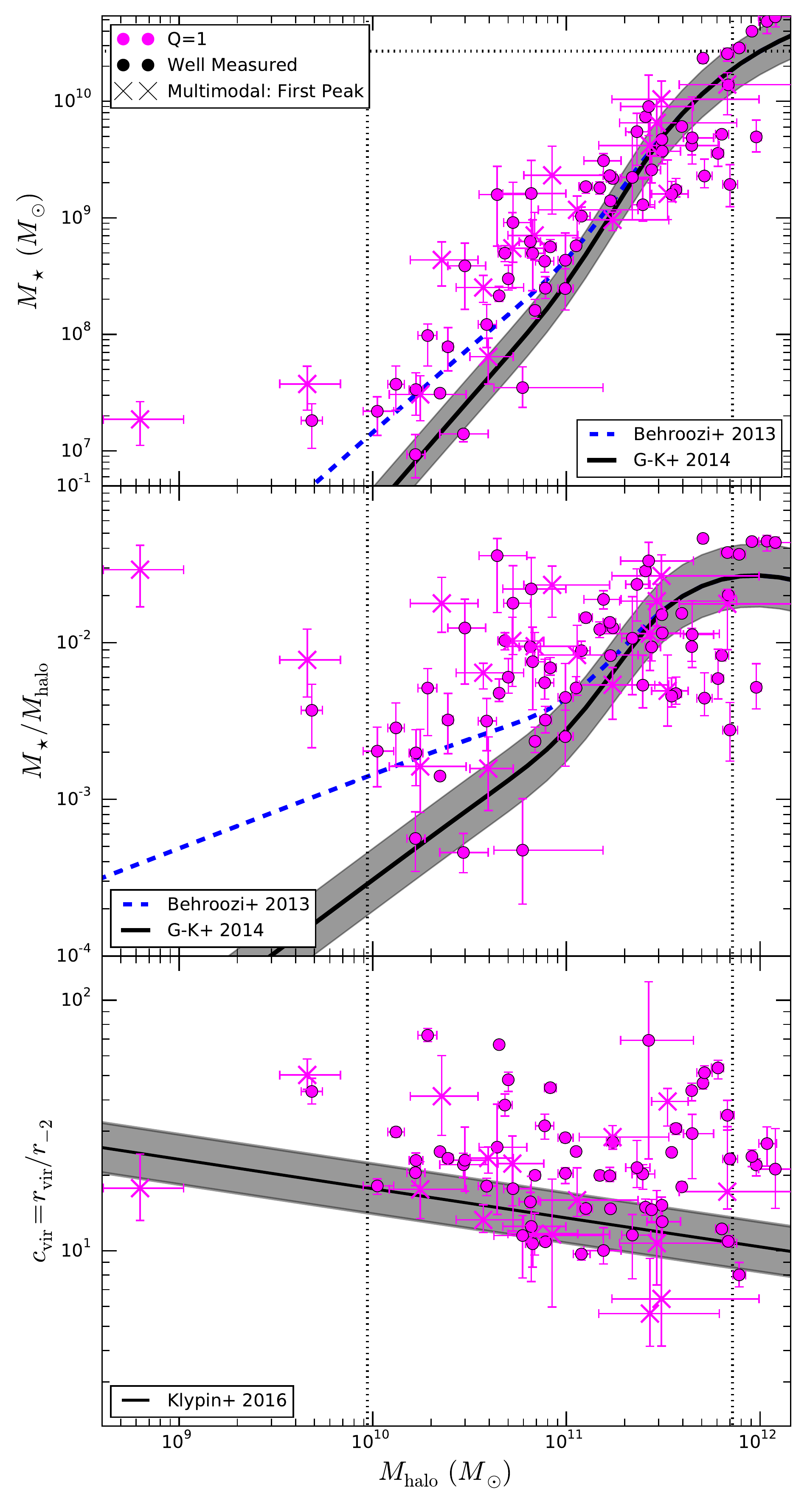}
\end{center}
\caption{Same as Figure~\ref{fig:mstar_mhalo} with only Q=1 subset.
 }
\label{fig:mstar_mhalo_q_1}
\end{figure}

In this section we consider the final sample with only the Q=1 subset.  
We reproduce Figure~\ref{fig:mstar_mhalo} with this subset in Figure~\ref{fig:mstar_mhalo_q_1}.
Our main results are robust to including the Q=2,3 subsets.

\section{Deviation from Cosmological Relationships}
\label{app:deviation}
\begin{figure*}
\begin{center}
\includegraphics[scale=.5]{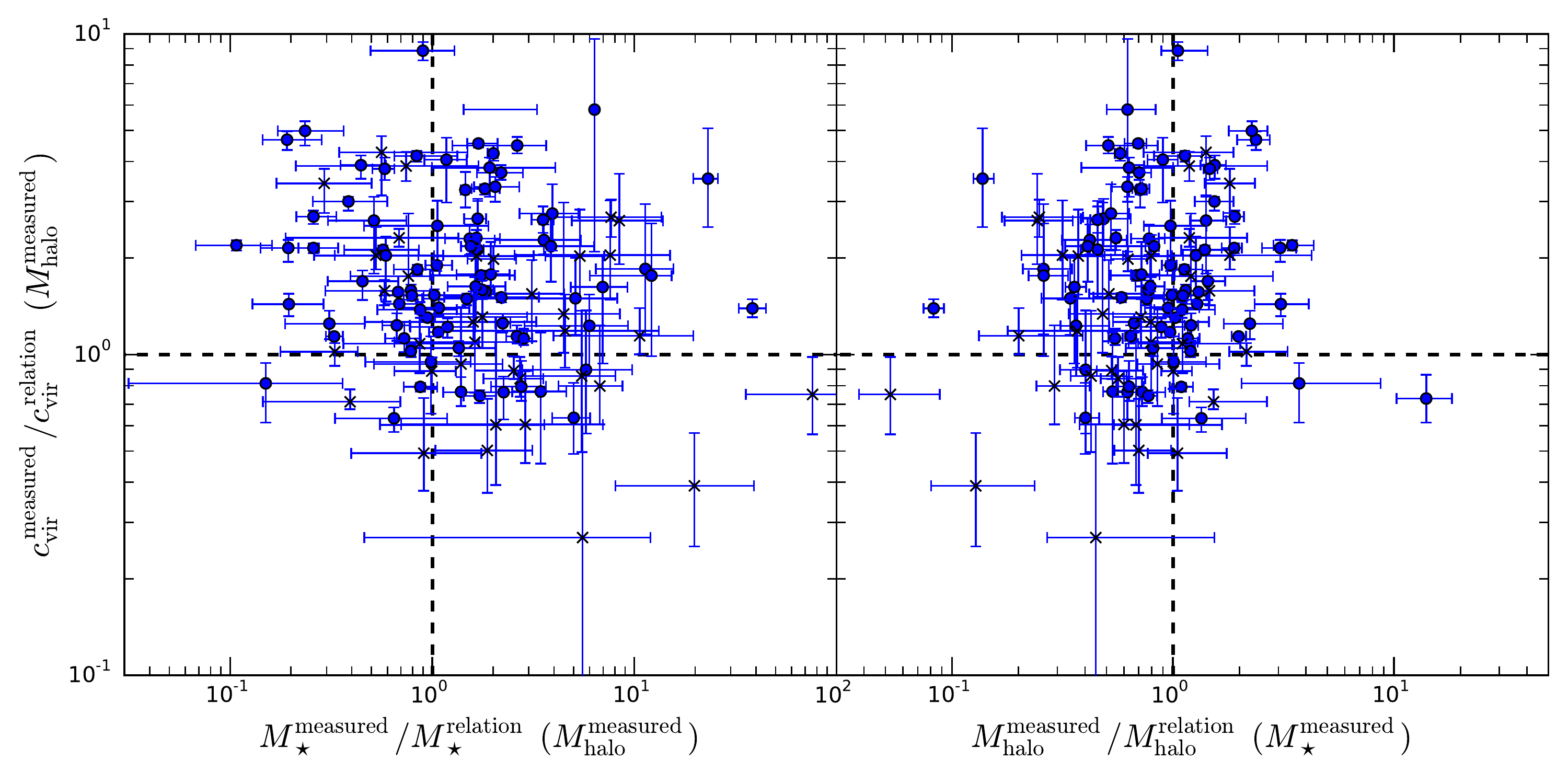}
\end{center}
\caption{Ratio of the deviation from the \citet{Behroozi2013_abundance} $\Mstar-\Mhalo$ relationship versus the deviation from the \citet{Klypin2016}  $\cvir-\Mhalo$ relationship.
The deviation from $\Mstar-\Mhalo$ relation is computed by comparing the ratio of the measured $\Mstar$ to the expected $\Mstar$ at the measured $\Mhalo$ (left panel) or comparing the ratio of  the measured $\Mhalo$ to the expected $\Mhalo$ for the measured $\Mstar$ (right panel).
The deviation from $\cvir-\Mhalo$ relation is computed by comparing the ratio of the measured $\cvir$ to the expected $\cvir$ at the measured $\Mhalo$.
Multimodal systems are shown as x's while single-mode systems are circles.
Dashed lines show where the measured value is equal to the relation.
 }
\label{fig:con_mhalo_deviation}
\end{figure*}

We explore the deviations from the \citet{Behroozi2013_abundance} $\Mstar-\Mhalo$ and \citet{Klypin2016}  $\cvir-\Mhalo$ relationships in this section.
In Figure~\ref{fig:con_mhalo_deviation}, we show the deviation from the $\Mstar-\Mhalo$ versus deviation from the $\cvir-\Mhalo$ relationship.  
Each deviation is expressed as the ratio of the measured quantity to the expected quantity at a fixed measurement.  
The $\cvir$ deviation is quantified by the ratio of the measured $\cvir$ to the expected $\cvir$ at the measured $\Mhalo$ value.  
We quantify the deviation from the $\Mstar-\Mhalo$ relationship two ways.  
First (left panel in Figure~\ref{fig:con_mhalo_deviation}), we compute the ratio of the measured $\Mstar$ versus the expected $\Mstar$ at the measured $\Mhalo$ value.
Second (right panel in Figure~\ref{fig:con_mhalo_deviation}), we compute the ratio of the measured $\Mhalo$ versus the expected $\Mhalo$ at the measured $\Mstar$ value.
In both cases, no trends between the deviations in the cosmological relationships are observed.

\section{Larger Kinematic $\Upsilon$}
\label{appendix}

\begin{figure*}
\begin{center}
\includegraphics[scale=.4]{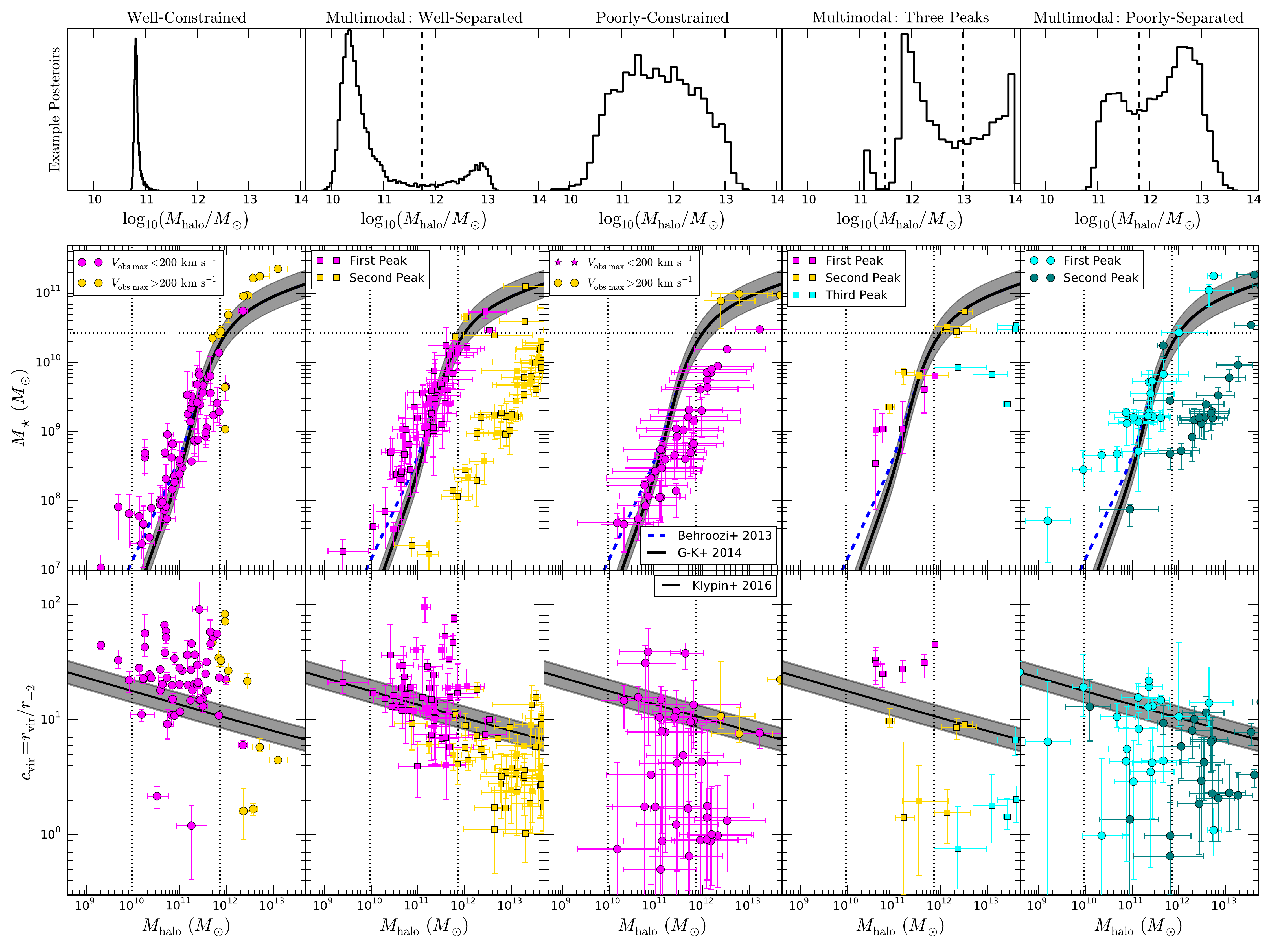}
\end{center}
\caption{ Same as Figure~\ref{fig:explain_cuts} but with a larger prior range on $\Upsilon_{{\rm kinematic }}$.
There is an additional column for multimodal systems with distinct modes in the $\Mhalo$ posterior.
 }
\label{fig:explain_cuts_large_ml}
\end{figure*}

\begin{figure}
\begin{center}
\includegraphics[scale=.425]{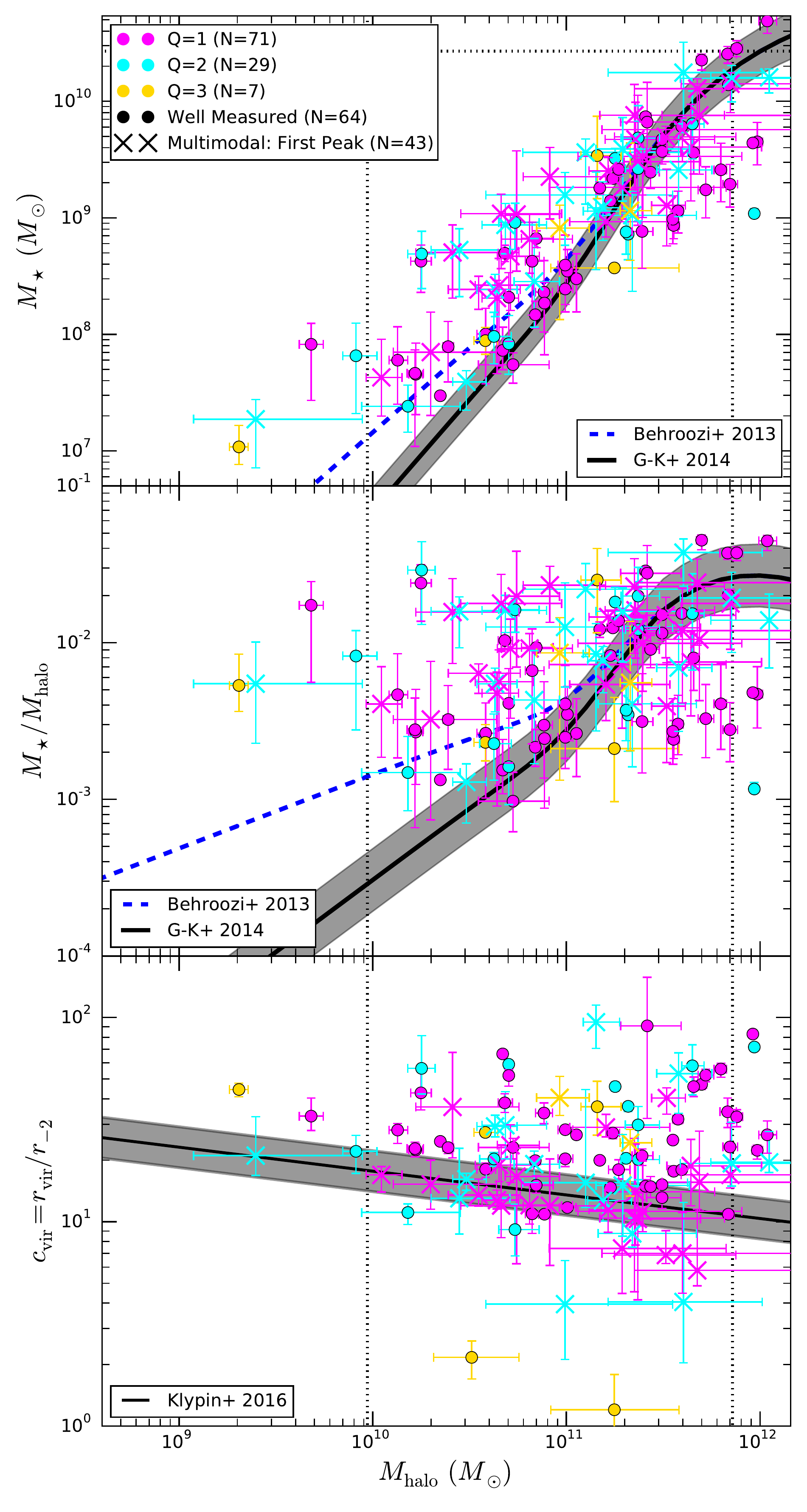}
\end{center}
\caption{Same as Figure~\ref{fig:mstar_mhalo} but with a larger prior range on $\Upsilon_{{\rm kinematic }}$.
 }
\label{fig:mstar_mhalo_large_ml}
\end{figure}

We explore a larger prior range in $\Upsilon_{{\rm kinematic}}$ is this section.  
The prior range is increased to $0.05< \Upsilon_{{\rm kinematic}} < 10.0$.
We reproduce Figures~\ref{fig:explain_cuts} and ~\ref{fig:mstar_mhalo} with the larger prior range in Figures~\ref{fig:explain_cuts_large_ml} and ~\ref{fig:mstar_mhalo_large_ml} respectively.
Our main results are robust to the increased prior range but the results for individual galaxies are not.  
There are more systems with multimodal $\Mhalo$ posteriors and several systems have three distinct modes in the $\Mhalo$ posterior. 
We include an additonal column in Figure~\ref{fig:explain_cuts_large_ml} to show the three mode $\Mhalo$ systems.

\bsp	
\label{lastpage}
\end{document}